\documentclass{emulateapj}

\slugcomment{Accepted for publication in the Astrophysical Journal}

\shorttitle{ROTATIONALLY RESOLVED C$_3$ }
\shortauthors{\'AD\'AMKOVICS ET AL.}

\usepackage{natbib}
\begin{document}

\title{Observations of Rotationally Resolved C$_3$ \\ in Translucent Sight Lines}

\author{M\'at\'e \'Ad\'amkovics,\altaffilmark{1} Geoffrey A. Blake,\altaffilmark{2}
and Benjamin J. McCall\altaffilmark{1,3}}

\altaffiltext{1}{Department of Chemistry, University of
California, Berkeley, CA, 94720; mate@haze.cchem.berkeley.edu}
\altaffiltext{2}{Division of Geological and Planetary Sciences and
Division of Chemistry and Chemical Engineering, M/S 150-21,
California Institute of Technology, Pasadena, CA 91125.}
\altaffiltext{3}{Astronomy Department, 601 Campbell Hall,
University of California, Berkeley, CA 94720}

%%%%%%%%    ABSTRACT   %%%%%%%%%%%%%%%%%%%%%%%%%%%%%%%%%%%%%%%%%%%

\begin{abstract}
The rotationally resolved spectrum of the $A ^1\Pi_u \leftarrow X
^1\Sigma^+_g$ 000-000 transition of C$_3$, centered at 4051.6~{\AA}, has been
observed along 10 translucent lines of sight. To
interpret these spectra, a new method for the determination of
column densities and analysis of excitation profiles involving the
simulation and fitting of observed spectra has been developed. The
populations of lower rotational levels ($J$$\le$14) in C$_3$ are
best fit by thermal distributions that are consistent with the kinetic
temperatures determined from the excitation profile of C$_2$.
Just as in the case of C$_2$, higher rotational levels
($J$$>$14) of C$_3$ show increased nonthermal population distributions in
clouds which have been determined to have total gas
densities below $\sim$500 cm$^{-3}$.
\end{abstract}

%%%%%%%%    ABSTRACT


\section{Introduction} \label{intro}

The existence of carbon chain molecules in diffuse interstellar
clouds raises questions about the chemistry that occurs in
these clouds as well as the fate of large carbonaceous
molecules produced throughout the life-cycles of stars. Carbon
chain molecules have been suggested as the source of the
confounding diffuse interstellar bands \citep{Douglas1977},
although efforts to identify a specific molecular carrier have so far
failed. C$_7^-$  and l-C$_3$H$_2^-$ have been the most recently
rejected 
\citep{McCall2001, McCall2002b}
in a long list of
candidates \citep{Herbig1995}. The smallest carbon chain molecule,
C$_2$, has been used as a powerful probe of the temperature and
density in diffuse clouds \citep{vD&B1989} and, until very
recently, diffuse cloud chemistry was understood only in terms of
atoms and diatomics.

After repeated efforts 
\citep{Clegg1982, Snow1988, Haffner1995}, the
first clear detections of C$_3$ in three diffuse clouds were
reported by \citet{Maier2001}.  Since this first observational
study of C$_3$ in the diffuse interstellar medium, a detailed model has
been developed to understand its excitation \citep{Roueff2002} and
a low resolution survey has detected it in 15 translucent sight
lines \citep{Oka2003}. Along with infrared measurements of H$_3^+$
\citep{McCall1998,McCall2002a,Geballe1999}, which can directly
probe the interstellar cosmic ray ionization rate \citep{McCall2003a},
the detections of C$_3$ demonstrate that the chemistry of diffuse 
clouds may be significantly richer than expected. 

The ``$\lambda$4050 group'' of C$_3$ was first observed in emission
in comet Tebutt \citep{Huggins1881}, and has since been observed
in the photospheres of many cool stars \citep{Jorg1994}, in a few
protoplanetary nebulae \citep{Hriv1999}, and toward many other
comets (see, for example, recent work by \citet{Rouss2001}).
The $\lambda$4050 group was first observed in the laboratory
by C. W. \citet{Raffety1916} in emission from the flame of a Mecker
burner. \citet{Douglas1951} identified the carrier of the group as C$_3$,
and a detailed assignment as $A ^1\Pi_u \leftarrow X ^1\Sigma^+_g$ was 
made by \citet{Gausset1965}. At longer wavelengths, C$_3$ has also 
been observed in the material around the asymptotic giant branch star 
IRC +10216 using both the asymmetric vibrational stretching mode at 
4.9$\mu$m \citep{Hinkle1988} and the anomalously low frequency bending 
mode at 157$\mu$m \citep{Cerni2000}. 

As a linear carbon chain, like C$_2$, C$_3$ has no permanent dipole
moment and thus lacks the ability to cool efficiently via rotational
transitions. High rotational levels of C$_3$, once produced during 
chemical formation or following UV pumping, can only relax slowly by 
collisional de-excitation and have significant lifetimes under 
interstellar conditions. \citet{Maier2001} reported that the low 
$J$ levels ($J$$<$14) in their spectra of C$_3$ toward $\zeta$ Oph 
were fit by a Boltzmann distribution at $T_l$=60 K, whereas $J$$>$14 
were fit by a Boltzmann distribution at $T_h$=230 K. 

%%%%%%%%%%%%%%%%%%%%%%%%%%%%%%%%%%%%%%%%%%%%%%%%%%%%%%%%%%%%%%%%%%%%%%%%%%%%%%%%%%
%%
%%   Program Stars Table
%%
%%%%%%%%%%%%%%%%%%%%%%%%%%%%%%%%%%%%%%%%%%%%%%%%%%%%%%%%%%%%%%%%%%%%%%%%%%%%%%%%%%

\begin{deluxetable*}{cccccccc}
\tablecolumns{8}
\tabletypesize{\footnotesize}
\tablewidth{0pc}
\tablecaption{Data For Program Stars \label{tbl-prog}}
\tablehead{\colhead{Star}                 & \colhead{Name}   &
\colhead{Type}                 & \colhead{\em{V}} &
\colhead{{\em E(B-V)}}         & \colhead{\em{D} (pc)\tablenotemark{a}}   &
\colhead{S/N\tablenotemark{b}} & \colhead{Date (UT)}}
\startdata

HD 11415\phn   & $\epsilon$ Cas & B3 III & 3.38 & 0.05  & 136 & 2400 &2002 Dec 15 \\
HD 20041\phn   &\nodata    & A0 Ia    & 5.79 & 0.72 & 440  & 1000 & 2002 Dec 14 \\
HD 21389\phn   &\nodata    & A0 Iae   & 4.54 & 0.57 & 940  & 1300 & 2002 Dec 14 \\
HD 21483\phn   &\nodata    & B3 III   & 7.06 & 0.56 & 440  & 1000 & 2002 Dec 15 \\
HD 22951\phn   &40 Per     & B0.5 V   & 4.97 & 0.27 & 283  & 1700 & 2002 Dec 15 \\
HD 23180\phn   &$o$ Per    & B1 III   & 3.83 & 0.31 & 280  & 1850 & 2002 Dec 14 \\
HD 24398\phn   &$\zeta$ Per& B1 Ib    & 2.85 & 0.31 & 301  & 2200 & 2002 Dec 15 \\
HD 24534\phn   &X Per      & O9.5pe   & 6.10 & 0.59 & 590  & 1200 & 2002 Dec 15 \\
HD 24912\phn   &$\xi$ Per  & O7e      & 4.04 & 0.33 & 470  & 2600 & 2002 Dec 14 \\
HD 34078\phn   &AE Aur     & O9.5 Ve  & 5.96 & 0.52 & 620  & 1200 & 2002 Dec 14 \\
HD 35149\phn   &23 Ori     & B1 V     & 5.00 & 0.11 & 450  & 2100 & 2002 Dec 15 \\
HD 36371\phn   &$\chi$ Aur & B5I ab   & 4.76 & 0.43 & 880  & 2000 & 2002 Dec 15 \\
HD 37022\phn   &$\theta^1$ Ori C & O6 & 5.13 & 0.34 & 450  & 2000 & 2002 Dec 15 \\
HD 37061\phn   &\nodata    & B1 V     & 6.83 & 0.52 & 580  & 1300 & 2002 Dec 15 \\
HD 41117\phn  &$\chi^2$ Ori  &B2 Iae & 4.63 & 0.45 & 1000 & 2100 & 2002 Dec 15 \\
HD 42087\phn   &3 Gem      & B.25 Ibe & 5.75 & 0.36 & 1200 & 1600 & 2002 Dec 14 \\
HD 43384\phn   &9 Gem      & B3 Ib    & 6.25 & 0.58 & 1100 & 1600 & 2002 Dec 14 \\
HD 53367\phn   &\nodata    & B0 IVe   & 6.96 & 0.74 & 780  & 1300 & 2002 Dec 15 \\
HD 62542\phn   &\nodata    & B5 V     & 8.04 & 0.35 & 246  & 700  & 2002 Dec 14 \\
HD 169454  &\nodata    & B1.5 Ia  & 6.61 & 1.12 & 930  & 450  & 2002 Jul 14 \\
HD 204827  &\nodata    & B0 V     & 7.94 & 1.11 & 600  & 650  & 2002 Dec 14, 2002 Jul 13 \\
HD 206267  &\nodata    & O6 f     & 5.62 & 0.53 & 1000 & 1200 & 2002 Jul 12, 2002 Jul 13 \\
HD 207198  &\nodata    & O9 IIe   & 5.95 & 0.62 & 1000 & 700  & 2002 Jul 12, 2002 Jul 13 \\
HD 210839  &$\lambda$ Cep & O6 If & 5.04 & 0.57 & 505  & 700  & 2002 Jul 13 \\
\enddata
\tablenotetext{a}{The distance to $\theta^1$ Ori C is from 
Shuping and Snow (1997). All other distances are taken from  
Thorburn et al. (2003) and Oka et al. (2003).}
\tablenotetext{b}{Signal to noise per pixel, measured near 4049~{\AA}.}
\end{deluxetable*}

%%%%%%%%%%%%%%%%%%%%%%%%%%%%%%%%%%%%%%%%%%%%%%%%%%%%%%%%%%%%%%%%%%%%%%%%%%%%%%%%%

Depending on the intensity of the UV radiation field, 
the rates of collisional (de-)excitation and UV pumping,
and the photochemical lifetime of C$_3$, \citet{Roueff2002} suggest that 
the high rotational levels of C$_3$ may retain information about the
temperature at which C$_3$ was formed as well as the temperature
and density of the cloud.
\citet{Roueff2002} applied their sophisticated C$_3$ excitation model 
to all three of the detections by \citet{Maier2001} -- $\zeta$ Oph, 
$\zeta$ Per, and 20 Aql -- along with their own observations of
HD 210121, but none of the observations determined the population
of C$_3$ in $J>14$ well enough to constrain the full excitation model.
\citet{Galazut2002} have since provided spectra of C$_3$ toward $\chi$ 
Oph and HD 152236, without analysis of the excitation.

With the guidance of the low resolution survey results of
\citet{Oka2003} we have measured the spectra of 24 reddened stars
at high resolution in order to study the excitation profile of C$_3$ 
in a large sample of diffuse clouds. Here we present the rotationally
resolved spectra of C$_3$ at high signal to noise (S/N) 
in 10 of these translucent sight lines. We present a new spectral 
fitting method for determining the populations in each $J$ level 
and examine the relationship between the rotational excitation of 
C$_3$ and C$_2$.

\section{Observations} \label{obs}

Observations were carried out using the Shane 3-m telescope and
the Hamilton Echelle Spectrograph \citep{Vogt1987} at Lick
Observatory on 2002 July 12 through 14 and using the High
Resolution Echelle Spectrometer (HIRES) 
\citep{Vogt1994}
on the
10-m Keck I telescope atop Mauna Kea on 2002 December 14 and 15.

The Hamilton spectrograph operates with an echelle grating and two
cross-dispersing prisms, is mounted at coud\'e focus, and provides
nearly complete spectral coverage from 3,500-10,000 {\AA}. The
detector used was the Lick3 L4-9-00AR thinned 2048$\times$2048 CCD with
15$\mu$m pixels. The 1.$\arcsec$2 wide and 2.$\arcsec$0 long slit
provides a resolving power $R$=$\lambda/\Delta\lambda$$\sim$60,000,
where the resolution $\Delta\lambda$ refers to the full width at
half maximum (FWHM) of the instrument profile, $\sim$0.1 {\AA} at
6,000 {\AA}, which spans two pixels on the detector. The reduced data 
have a dispersion $d$=0.0336 {\AA} pixel$^{-1}$ at 4050 {\AA}.

The HIRES instrument is also a cross-dispersed echelle
spectrograph, and features a backside-illuminated Tek 2048x2048
CCD detector functioning over the 3,500-10,000 {\AA} range and
capturing a spectral span of 1200-2500 {\AA} per exposure. To
minimize total readout time for our relatively bright targets,
high gain mode and dual-amp readout were used. For these
observations spectra in the 3985-6420 {\AA} range were recorded
with each exposure, with small gaps in the spectral coverage
starting at $\sim$5000 {\AA} and increasing to $\sim$30 {\AA} at
6315 {\AA}, due to echelle orders falling off the edge of the CCD.
A slit size of 0.$\arcsec$574 $\times$ 14$\arcsec$ was used to
achieve a resolving power of $R$$\sim$67,000. The detector
position was constant throughout each night and moved only
slightly between the two nights of observations. The reduced data have
a dispersion $d$=0.0287 {\AA} pixel$^{-1}$ at 4050 {\AA}.

The data were reduced using standard procedures
in the echelle package in NOAO's Image Reduction and Analysis
Facility (IRAF). As many as 120 flat field images were taken
before and 180 after the observations to ensure that the S/N
of the reduced spectra was photon-counting limited. Calibration
and target images were bias corrected and combined, before target 
images were flat fielded and echelle apertures extracted. 
Wavelength calibration was performed with standard routines using 
spectra of a ThAr lamp, and one-dimensional spectra of the targets 
were extracted for further analysis. Generally, the entire
aperture blaze was fit with a low order polynomial to normalize
the spectra. In cases where the spectral features of interest were
observed very close to the center (maximum) of the aperture blaze,
only a small spectral region around the features was continuum fit
with a low order polynomial. The program stars and their
properties, along with the S/N achieved at 4049 {\AA}, are shown in
Table \ref{tbl-prog}.  Figure 1 presents the C$_3$ spectra
observed along the 10 lines of sight with detectable C$_3$
absorption.

%%%%%%%%%%%%%%%%%%%%%%%%%%%%%%%%%%%%%%%%%%%%%%%%%%%%%%%%%%%%%%%%%%%%%%%%%%%%%%%%%%
%%
%%   C3 Molecular Data Table
%%
%%%%%%%%%%%%%%%%%%%%%%%%%%%%%%%%%%%%%%%%%%%%%%%%%%%%%%%%%%%%%%%%%%%%%%%%%%%%%%%%%

\begin{deluxetable}{ccc}
\tabletypesize{\tiny}
\tablecolumns{3}
\tablewidth{0pc}
\tablecaption{Molecular Data for the $A ^1\Pi_u \leftarrow X ^1\Sigma^+_g$
transition of C$_3$ \label{tbl-C3moldata}}
\tablehead{
\colhead{Wavelength (\AA)} & \colhead{Line}  &
\colhead{$f_{J',J''}$ ($\times 10^3$)\tablenotemark{a}}}
\startdata
4049.770 & R(24) &  4.24 \\
4049.784 & R(26) &  4.23 \\
4049.784 & R(22) &  4.26 \\
4049.810 & R(20) &  4.30 \\
4049.810 & R(28) &  4.21 \\
4049.861 & R(18) &  4.33 \\
4049.861 & R(30) &  4.20 \\
4049.963 & R(16) &  4.36 \\
4050.081 & R(14) &  4.42 \\
4050.206 & R(12) &  4.48 \\
4050.337 & R(10) &  4.57 \\
4050.495 &  R(8) &  4.70 \\
4050.670 &  R(6) &  4.92 \\
4050.866 &  R(4) &  5.34 \\
4051.069 &  R(2) &  6.40 \\
4051.250\tablenotemark{b} &       R(0) &      16.00 \\
(4051.309) &            &            \\
4051.461 &  Q(2) &  8.00 \\
4051.519 &  Q(4) &  8.00 \\
4051.590 &  Q(6) &  8.00 \\
4051.681 &  Q(8) &  8.00 \\
4051.741\tablenotemark{c} &  P(2) &  1.60 \\
4051.793 & Q(10) &  8.00 \\
4051.929 & Q(12) &  8.00 \\
4052.062 &  P(4) &  2.66 \\
4052.089 & Q(14) &  8.00 \\
4052.271 & Q(16) &  8.00 \\
4052.424 &  P(6) &  3.08 \\
4052.473 & Q(18) &  8.00 \\
4052.698 & Q(20) &  8.00 \\
4052.792 &  P(8) &  3.29 \\
4052.940 & Q(22) &  8.00 \\
4053.180 & P(10) &  3.43 \\
4053.208 & Q(24) &  8.00 \\
4053.490 & Q(26) &  8.00 \\
4053.591 & P(12) &  3.52 \\
4053.794 & Q(28) &  8.00 \\
4054.020 & P(14) &  3.58 \\
4054.112 & Q(30) &  8.00 \\
4054.458 & P(16) &  3.64 \\
4054.908 & P(18) &  3.67 \\
4055.373 & P(20) &  3.70 \\
4055.877 & P(22) &  3.73 \\
4056.410 & P(24) &  3.75 \\
4056.961 & P(26) &  3.77 \\
4057.531 & P(28) &  3.78 \\
4058.121 & P(30) &  3.80 \\
\enddata
\tablenotetext{a}{Here $f=$0.0160 was used, consistent with
\citet{Maier2001} and \citet{Oka2003}; \citet{Roueff2002} and \citet{Galazut2002}
use $f=$0.0146.}
\tablenotetext{b}{The $R(0)$ assignment by \citet{Gausset1965} (shown parenthetically)
is inconsistent with our observations. We use the observed value,
confirmed by laboratory measurements \citep{McCall2003b}, in the rotational 
level population model.}
\tablenotetext{c}{Calculated transition wavelength \citep{Roueff2002}.}
\end{deluxetable}

%%%%%%%%%%%%%%%%%%%%%%%%%%%%%%%%%%%%%%%%%%%%%%%%%%%%%%%%%%%%%%%%%%%%%%%%%%%%%%%%%%

\section{Spectrum of C$_3$} \label{spectra}

Unlike molecules which can radiate away excess energy, the
distribution of population among rotational levels for molecules
in the ISM without a permanent dipole moment, such as C$_3$, is a
delicate balance between collisional and radiative processes. This
has been studied in detail for the homonuclear diatomics H$_2$
(\cite{B&D1977}, and references therein) and C$_2$
\citep{vD&B1982}, and more recently a similar method of analysis
has been used to examine C$_3$ \citep{Roueff2002}. In general, these
models show that the populations of lower rotational levels
($J$$\lesssim$14 for C$_3$) are controlled mostly by the kinetic
temperature of the gas along the line of sight, whereas the
populations in higher rotational levels are determined by the
competition between radiative pumping and collisional de-excitation.
The net result can be a non-thermal enhancement of high $J$
populations in low density environments as observed by 
\citet{Maier2001} toward $\zeta$ Oph.

In order to determine the total column density of C$_3$ and
measure the population distribution among rotational levels, we
have used two models to fit 40 rovibronic transitions between
4049--4055 {\AA}. A thermal excitation model incorporating either
one or two temperatures was used, as well as a model that fit the
population in each rotational level to the observed spectra. These
two methods provided a means to extract information from
overlapping or unresolved transitions in the C$_3$ $Q$-branch and
$R$-branch bandhead.

Laboratory measurements \citep{Gausset1965} of transition wavelengths
were used in both models, with the exception of a shifted $R(0)$ line 
that was used in the rotational level population model 
(see Section \ref{R0}).

%%%%%%%%%%%%%%%%%%%%%%%%%%%%%%%%%%%%%%%%%%%%%%%%%%%%%%%%%%%%%%%%%%%%%%%%%%%%%%%
%%
%%   All Detections and Thermal Model
%%
%%%%%%%%%%%%%%%%%%%%%%%%%%%%%%%%%%%%%%%%%%%%%%%%%%%%%%%%%%%%%%%%%%%%%%%%%%%%%%%

\begin{figure*}[t]
\epsscale{1.1} 
\plotone{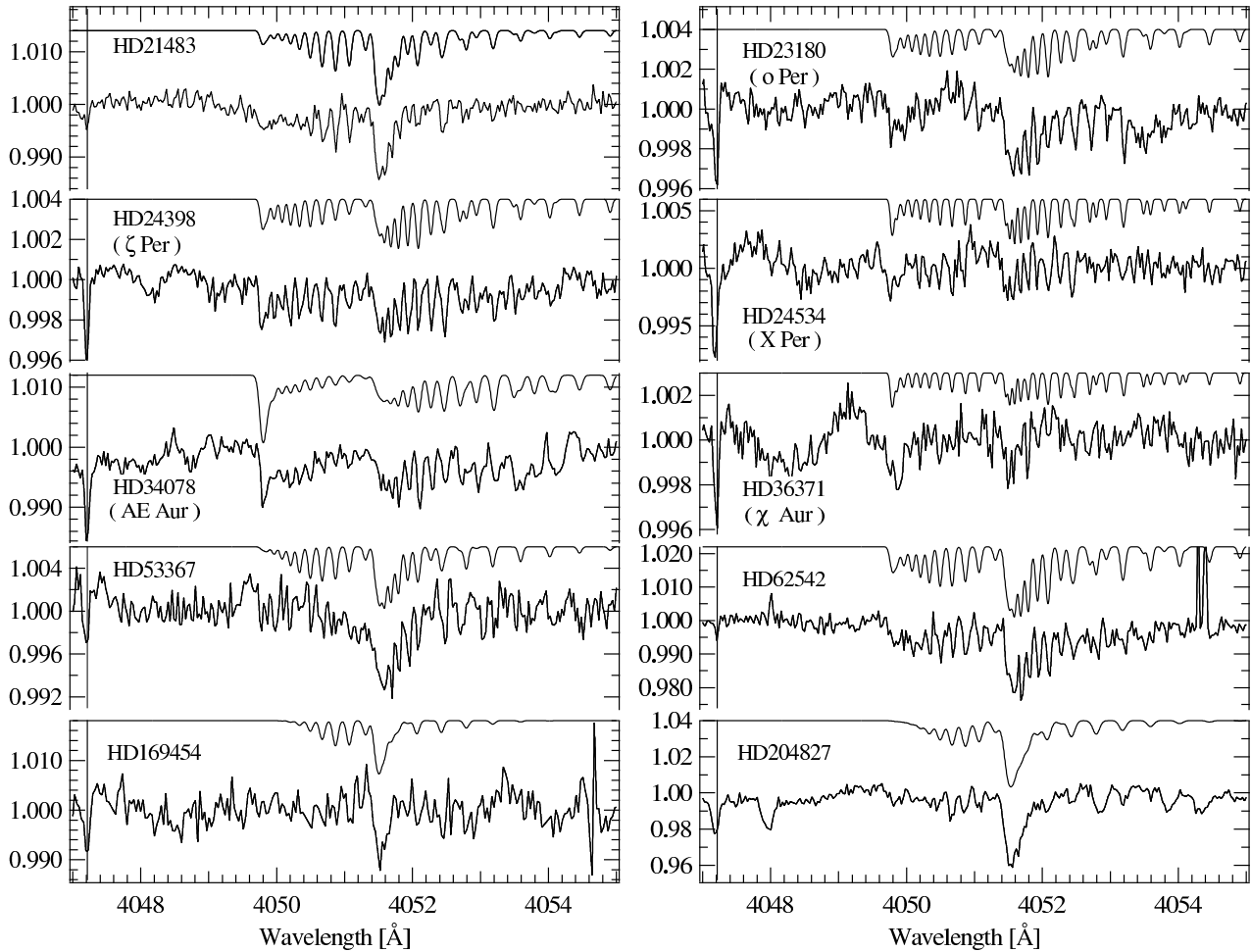} \caption{Spectra of
translucent sight lines toward which C$_3$ was detected, along with
thermal excitation model fits. Note how the $Q$-branch at
4051.6 {\AA} is very well fit, while the $R$ bandhead at
4049.8 {\AA} is sometimes underestimated. Spectra are all shifted
to rest wavelengths and the {\sc K i} line at 4047.2 {\AA} is
marked by a vertical line. Broad stellar lines have been removed
from the spectra of $\zeta$ Per, HD 21483, AE Aur, HD 36371, and
HD 53367 using Gaussian line profiles. \label{alldet}}
\end{figure*}

%%%%%%%%%%%%%%%%%%%%%%%%%%%%%%%%%%%%%%%%%%%%%%%%%%%%%%%%%%%%%%%%%%%%%%%%%%%%%%%%%%
%%
%%   Thermal (Boltzmann) Distribution Results
%%
%%%%%%%%%%%%%%%%%%%%%%%%%%%%%%%%%%%%%%%%%%%%%%%%%%%%%%%%%%%%%%%%%%%%%%%%%%%%%%%%%%

\begin{deluxetable*}{ccccccc}
\tablecolumns{7} 
\tabletypesize{\footnotesize}
\tablewidth{0pc}
\tablecaption{Best Fit Parameters for Thermal
Excitation Model of the Observed Spectra\label{tbl-Bmod}}
\tablehead{
\colhead{Star} & \colhead{$N$(C$_3$)\tablenotemark{a}}
& \colhead{$N_l$(C$_3$)\tablenotemark{a}} &
\colhead{$N_h$(C$_3$)\tablenotemark{a}} & \colhead{$T_l$} &
\colhead{$T_h$} & \colhead{FWHM\tablenotemark{b}}         \\
& & & & & & \colhead{(km s$^{-1}$)}
}
\startdata
HD 21483\phn & 4.60 $\pm$ 0.28 & 2.60    &   2.00  &   28.3 $\pm$  1.3 & 150 $\pm$ 13   &  5.6 $\pm$ 0.1 \\
HD 23180\phn &   1.27 $\pm$ 0.13 & \nodata & \nodata &  127.3 $\pm$  3.9 & \nodata      &  5.2 $\pm$ 0.2 \\
HD 24398\phn &   1.30 $\pm$ 0.35 & \nodata & \nodata &    132 $\pm$ 10.3 & \nodata      &  5.1 $\pm$ 0.4 \\
HD 24534\phn &   1.77 $\pm$ 0.19 &   0.49  &   1.28  &     40 $\pm$  6.1 & 250 $\pm$ 41 &  3.7 $\pm$ 1.0 \\
HD 34078\phn &   6.21 $\pm$ 0.68 & \nodata & \nodata &  301.4 $\pm$ 15.4 & \nodata      &  7.2 $\pm$ 0.3 \\
HD 36371\phn &   0.75 $\pm$ 0.06 &   0.13  &   0.62  &   20.1 $\pm$    2 & 250 $\pm$ 52 &  3.1 $\pm$ 1.2 \\
HD 53367\phn &   2.08 $\pm$ 0.26 & \nodata & \nodata &   60.8 $\pm$  1.5 & \nodata      &  5.6 $\pm$ 0.1 \\
HD 62542\phn &  10.37 $\pm$ 0.53 &   7.35  &   3.01  &   75.5 $\pm$  7.2 & 200 $\pm$ 76 &  5.6 $\pm$ 0.1 \\
HD 169454    &   2.24 $\pm$ 0.66 & \nodata & \nodata &   23.4 $\pm$  1.4 & \nodata      &  6.5 $\pm$ 0.4 \\
HD 204827    &  11.13 $\pm$ 0.87 & \nodata & \nodata &   40.6 $\pm$  0.8 & \nodata      &  8.5 $\pm$ 0.2 \\
\enddata
\tablenotetext{a}{Column density $N$ in 10$^{12}$ cm$^{-2}$.}
\tablenotetext{b}{Full-width at half maximum derived from direct fit to the 
observed spectrum; instrumental line width at 4050 {\AA} is $\sim$4.5 km s$^{-1}$.}
\end{deluxetable*}

%%%%%%%%%%%%%%%%%%%%%%%%%%%%%%%%%%%%%%%%%%%%%%%%%%%%%%%%%%%%%%%%%%%%%%%%%%%%%%%

\subsection{Thermal Excitation Model} \label{thermal}

In this model the relative population of each rotational level 
($J$$\le$30) is
determined using a thermal distribution at temperature, $T$. The
fraction of total molecules in a particular $J$ level is given by
\begin{equation}
F_J(T) = \frac{2J+1}{q_r}exp\left(\frac{-hcB_0}{kT}J(J+1)\right)
\end{equation}
where $q_r$ is the rotational partition function given by
$$
q_r = \sum_{J\: even}^{\infty}(2J+1)e^{-hcB_0J(J+1)/kT}
$$
For C$_3$ the ground state rotational constant $B_0$=0.43057 
cm$^{-1}$ \citep{Schmutt1990}. 
The two parameters $T$ and $N$(C$_3$), along with the
$f$-values shown in Table \ref{tbl-C3moldata}, then determine the
equivalent width, $W_{J',J''}$, of each transition using the
standard relationship
\begin{equation}
W_{J',J''} = 8.853\times10^{-21} \lambda^2 N_{J} f_{J',J''}\;\mathrm{\AA}
\end{equation}
where $f_{J',J''}$, the oscillator strength for a given transition $J'
\leftarrow J''$ at wavelength $\lambda$ in {\AA}, is given by the
product of the electronic oscillator strength $f_{ul}$, the
Franck-Condon factor, and the H\"{o}nl-London factor. The units
for $N_J$ and $W_J$ are cm$^{-2}$ and {\AA}, respectively. Here we
use $f$=0.016, consistent with \citet{Maier2001} and
\citet{Oka2003}, for the product $f$ of the electronic
oscillator strength and Franck-Condon factor, whereas
\citet{Roueff2002} and \citet{Galazut2002} use $f$=0.0146. Future
determinations of the oscillator strength and Franck-Condon factor
will scale the derived column densities accordingly. The
H\"{o}nl-London factors for this band ($\Lambda = 1 \leftarrow 0$)
are $(J+2)/2(2J+1)$, 1/2, and $(J-1)/2(2J+1)$ for the $R$-, $Q$-,
and $P$-branch transitions, respectively. Table
\ref{tbl-C3moldata} shows the adopted wavelengths, assignments and 
$f_{J',J''}$ for each transition.

Beginning with an approximate estimate of the C$_3$ excitation
temperature ($T$$\sim$45 K), column density ($N$(C$_3$)$\sim$10$^{12}$
cm$^{-2}$), and line width (FWHM=5.2 km s$^{-1}$) a simulated spectrum was
produced.  Standard curve-fitting routines were then used to
minimize the difference between the observed and simulated
spectra, by varying these three parameters ($T$, $N$(C$_3$), and
FWHM). If a second temperature component was observed, in the form
of a pronounced $R$-branch bandhead that was not reproduced with
the single temperature simulation, then a spectrum incorporating
two temperatures $T_l$ and $T_h$, and column densities $N_l$(C$_3$) and
$N_h$(C$_3$) was used. 

The initial estimate for the line width was larger than the instrument
resolution ($\sim$4.5 km s$^{-1}$) and characteristic 
of the line widths in the observed spectra (Table \ref{tbl-Bmod}). 
Broad lines are observed because translucent sight lines 
typically sample multiple diffuse clouds, which can have 
velocity separations comparable to, or larger than,
our instrumental line width \citep{Welty2001}. 
The velocity distribution of the clouds, convolved with the instrumental 
line width, produces observed features that in some cases ({\em e.g.} HD 204827)
are rather wide. Since we are observing through multiple clouds, 
the temperatures derived from our model represent an integrated average over 
all the clouds along the line of sight. In the opposite extreme, 
the best fit line widths for X Per and $\chi$ Aur are slightly below the 
instrument profile due to the relatively large level of noise in those 
spectra. 

Figure \ref{alldet} shows the observed spectra
and best fit thermal excitation model for 10 detections, with
model results summarized in Table \ref{tbl-Bmod}. 
Measurement uncertainties are discussed in Section \ref{error}.

%%%%%%%%%%%%%%%%%%%%%%%%%%%%%%%%%%%%%%%%%%%%%%%%%%%%%%%%%%%%%%%%%%%%%%%%%%%%%%%
%%
%%   Boltzmann Plots and Rotational Level Fits
%%
%%%%%%%%%%%%%%%%%%%%%%%%%%%%%%%%%%%%%%%%%%%%%%%%%%%%%%%%%%%%%%%%%%%%%%%%%%%%%%%

\begin{figure*}
\epsscale{1} 
\plotone{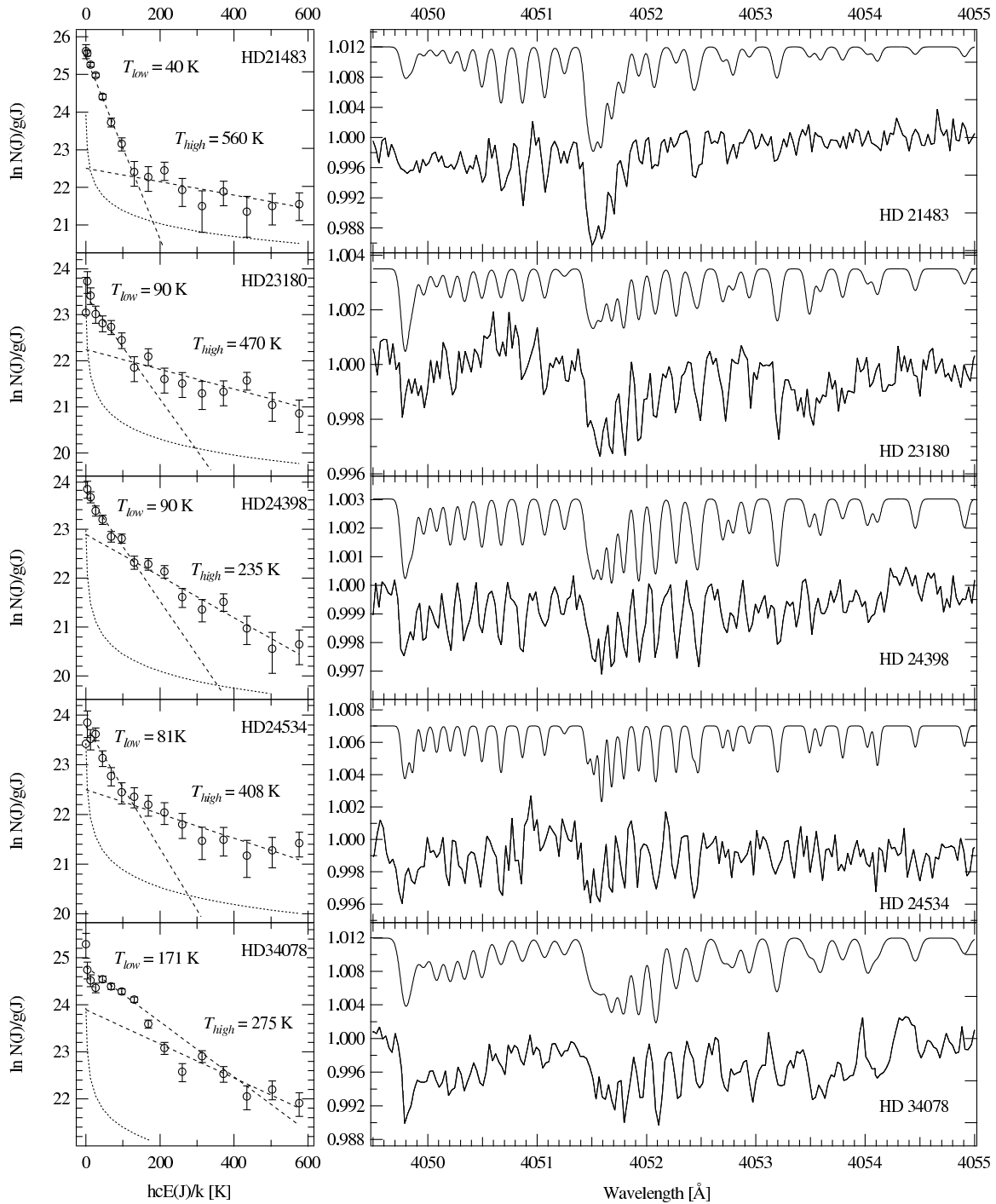} \caption{Spectra of C$_3$
detections with rotational excitation model fit (offset for clarity) 
and the Boltzmann plot ({\em left}) derived from fit to individual rotational 
levels. The dotted curve depicts the 1$\sigma$ detection limit as a function 
of $J$. Points without error bars indicate upper limits. Dashed lines indicate 
fits to datapoints for $J$$\le$14, and $J$$>$14, where the slope is inversely 
proportional to kinetic temperature. Note that the spectra are 
on different intensity scales and the very large columns of C$_3$ toward 
HD 62542 and HD 204827.  
\label{Bplots1}}
\end{figure*}

\begin{figure*}
\epsscale{1} 
\plotone{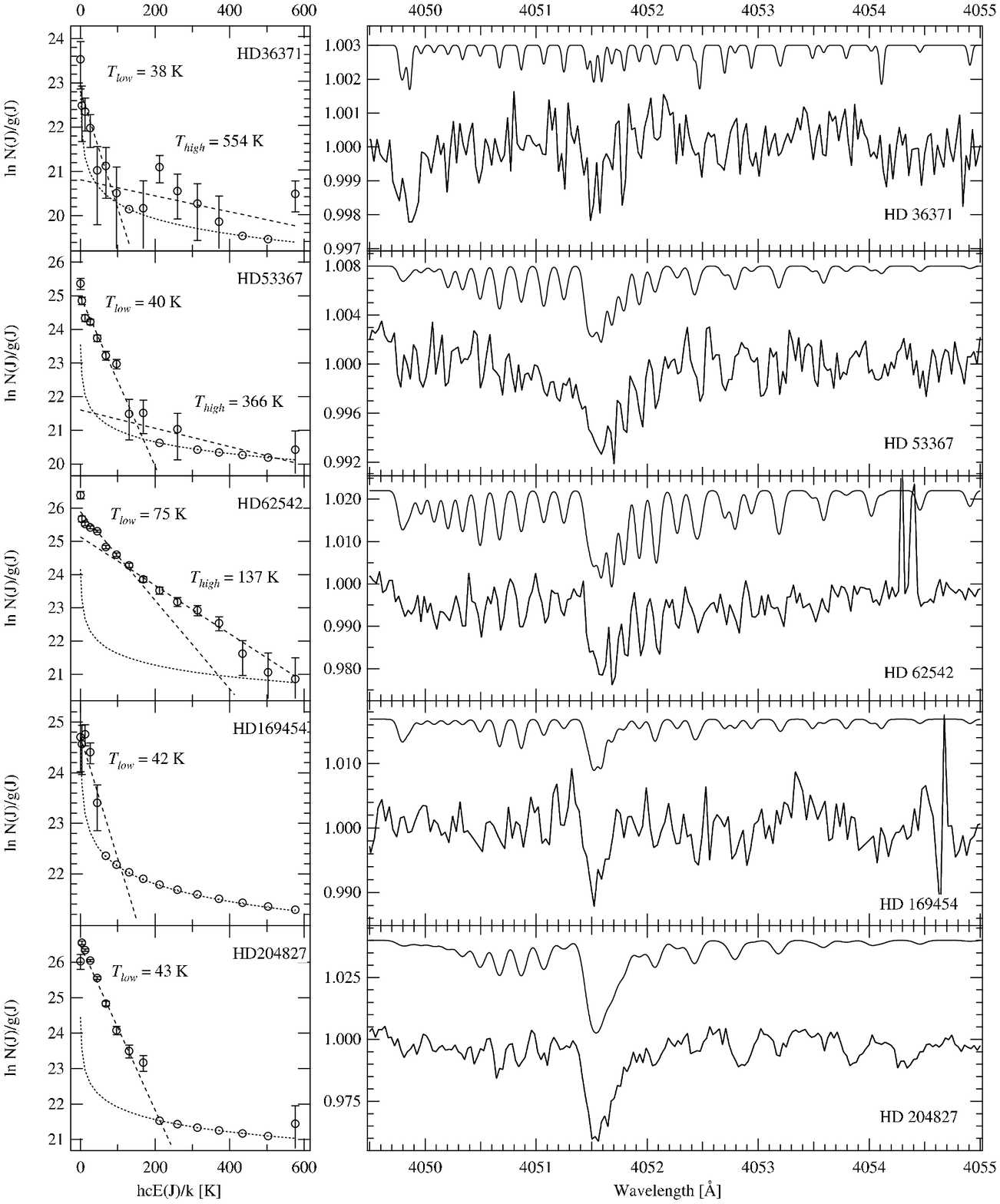}

Fig. 2.--- (continued)
\label{Bplots2}
\end{figure*}

\subsection{Rotational Level Population Model} \label{rotlev}

The most common method for determining the population of a
particular rotational level $J$ is to measure the equivalent
width $W$ of an unblended absorption feature. If multiple
transitions ($\Delta J=0,\pm1$) from the same lower $J$ state are
observed, then each transition probes the same population and
multiple measurements of the same quantity may be made.

In the case of our C$_3$ spectra, the low $J$ transitions
($J$$\lesssim$12) in the $R$-branch are well resolved, but the
stronger $Q$-branch lines starting from the same $J$ levels pile
atop one another. Similarly, transitions from higher $J$ levels that
are well resolved in the $Q$-branch ({\em e.g.} $Q$(16)  and $Q$(22)) are 
blended into a single feature in the $R$-branch bandhead. In order to
extract information from all of the observed transitions, a model
was used which varies the population in each $J$ level as a free
parameter, produces a simulated spectrum, and then minimizes the 
difference between the observed and simulated spectrum.

The initial population values for the model were determined from
the best fit thermal distribution. A curve fitting routine was
then used to vary each $N_J$ and minimize the residuals.
After fitting the population in each $J$ level, a Boltzmann plot 
was generated for each source,
and kinetic temperatures which best fit $J$$\le$14 and $J$$>$14 were
determined. Figure \ref{Bplots1} shows the observed and modeled
spectra, with Boltzmann plots derived from the fit. The detection 
limits for each $J$ are shown as dotted lines in the
Boltzmann plots. Results from the model are listed in Table
\ref{tbl-jmod} with uncertainties discussed below.

%%%%%%%%%%%%%%%%%%%%%%%%%%%%%%%%%%%%%%%%%%%%%%%%%%%%%%%%%%%%%%%%%%%%%%%%%%%%%%%
%%
%%   Rotational (J-) Level Excitation Results
%%
%%%%%%%%%%%%%%%%%%%%%%%%%%%%%%%%%%%%%%%%%%%%%%%%%%%%%%%%%%%%%%%%%%%%%%%%%%%%%%%

\begin{deluxetable*}{ccccccccccc}
\tablecolumns{11} \tabletypesize{\scriptsize} \tablewidth{0pc}
\tablecaption{Rotational Level Populations Determined by Fitting
the Observed Spectra\label{tbl-jmod}}
\tablehead{ & \multicolumn{10}{c}{$N_J$($\times 10^{11}$ cm$^{-2}$)} \\
\colhead{$J$}       &
\colhead{HD 21483}   &
\colhead{HD 23180}   &
\colhead{HD 24398}   &
\colhead{HD 24534}   &
\colhead{HD 34078}   &
\colhead{HD 36371}   &
\colhead{HD 53367}   &
\colhead{HD 62542}   &
\colhead{HD 169454}  &
\colhead{HD 204827}}
\startdata
 0 & 1.22 & 0.09 & 0.34 & 0.14 & 0.88 & 0.15 & 0.94 & 2.64 & 0.49 & 1.83 \\
 2 & 5.79 & 0.92 & 1.01 & 1.04 & 2.55 & 0.27 & 2.83 & 6.46 & 2.13 & 15.42 \\
 4 & 7.67 & 1.21 & 1.54 & 1.35 & 3.65 & 0.42 & 3.04 & 10.04 & 4.66 & 22.54 \\
 6 & 8.34 & 1.18 & 1.69 & 2.14 & 4.51 & 0.42 & 3.93 & 12.96 & 4.73 & 24.27 \\
 8 & 6.19 & 1.26 & 1.84 & 1.72 & 7.06 & 0.21 & 3.17 & 15.15 & 2.27 & 19.44 \\
10 & 3.87 & 1.43 & 1.61 & 1.49 & 7.53 & 0.29 & 2.34 & 11.59 & $<$ 0.96 & 11.68 \\
12 & 2.61 & 1.29 & 1.84 & 1.28 & 8.03 & 0.18 & 2.17 & 11.04 & $<$ 0.96 & 6.53 \\
14 & 1.43 & 0.82 & 1.30 & 1.36 & 7.85 & $<$ 0.15 & 0.57 & 9.22 & $<$ 0.96 & 4.24 \\
16 & 1.42 & 1.19 & 1.44 & 1.31 & 5.30 & 0.17 & 0.67 & 6.94 & $<$ 0.96 & 3.48 \\
18 & 1.90 & 0.82 & 1.39 & 1.26 & 3.57 & 0.49 & $<$ 0.31 & 5.54 & $<$ 0.96 & $<$ 0.76 \\
20 & 1.26 & 0.82 & 0.91 & 1.10 & 2.38 & 0.32 & 0.51 & 4.36 & $<$ 0.96 & $<$ 0.76 \\
22 & 0.90 & 0.73 & 0.78 & 0.87 & 3.65 & 0.26 & $<$ 0.31 & 3.73 & $<$ 0.96 & $<$ 0.76 \\
24 & 1.43 & 0.82 & 0.99 & 0.97 & 2.74 & 0.19 & $<$ 0.31 & 2.74 & $<$ 0.96 & $<$ 0.76 \\
26 & 0.91 & 1.13 & 0.62 & 0.76 & 1.83 & $<$ 0.15 & $<$ 0.31 & 1.19 & $<$ 0.96 & $<$ 0.76 \\
28 & 1.14 & 0.72 & 0.44 & 0.91 & 2.28 & $<$ 0.15 & $<$ 0.31 & 0.73 & $<$ 0.96 & $<$ 0.76 \\
30 & 1.28 & 0.64 & 0.52 & 1.12 & 1.83 & 0.44 & 0.42 & 0.64 & $<$ 0.96 & $<$ 1.15 \\
   &      &      &      &      &      &      &      &      &      &      \\
$\delta N_J$\tablenotemark{a}& 0.40 & 0.21 & 0.17 & 0.27 & 0.38 & 0.15 & 0.31 & 0.57 & 0.96 & 0.76 \\
\enddata
\tablenotetext{a}{The 1$\sigma$ uncertainty $\delta N_J$ is determined using 
Eqns 2 and 3 and $f$=0.016. Note that $\delta N_J$ for $R(0)$ is half of the
listed value.}
\end{deluxetable*}

%%%%%%%%%%%%%%%%%%%%%%%%%%%%%%%%%%%%%%%%%%%%%%%%%%%%%%%%%%%%%%%%%%%%%%%%%%%%%%%

\subsection{Uncertainties and Detection Limits} \label{error}

Uncertainties in the measurement of column densities for
individual rotational levels are dependent on the uncertainty in
the measurement of the equivalent width $W$. The 1$\sigma$
uncertainty in $W$ is given by
\begin{equation}
\delta W_{1\sigma} = (wd)^{1/2}\sigma
\end{equation}
where $w$ is the FWHM of the observed transition (in {\AA}), $d$ is
the dispersion of the instrument (in {\AA} pixel$^{-1}$), and
$\sigma$ is the standard deviation of the normalized
continuum--that is, (S/N)$^{-1}$. The $\delta W$ for each
transition is converted to an uncertainty in population, $\delta
N_J$ using Equation 2. For a given $J$ the transition with the
largest oscillator strength will give the smallest uncertainty in
$N_J$. In all cases except the $R(0)$ transition, the $Q$-branch
has the largest oscillator strength for a given $J$ (see Table
\ref{tbl-C3moldata}) and this is the $f$-value we use for our
uncertainty estimate.

The detection limits for the total column of C$_3$ are dependent
on the excitation of C$_3$. A larger total population of C$_3$ may
go undetected if it is distributed among many rotational levels
than if the population is all confined to a few low $J$ rotational
levels. For our upper limits listed in Table \ref{tbl-summary}
we assume that molecules are all distributed evenly among $J$=0 -- 20. 
Although the choice of 11 levels among which the distribution is 
evenly spread is not based on any underlying physics, a comparison 
with the spectrum of $\zeta$ Per shows that for the purposes of 
estimating our detection limit, such an approach is not unreasonable.

\subsection{The $R(0)$ Transition}\label{R0}

During the initial analysis of the Boltzmann plots it became
evident that $R(0)$ was uniformly underpopulated relative to what
was expected from a thermal distribution that fit $J$=2 -- 10. A
close inspection of the spectra revealed that the observed $R(0)$
transition was significantly shifted toward the blue from the
laboratory assignment of \citet{Gausset1965}. This was
particularly apparent in the spectrum of HD 62542, and can be noted
by comparing the thermal excitation model (which uses the
\citet{Gausset1965}  assignment of $R(0)$) with our observations
in Figure \ref{alldet}. Analysis of the higher resolution spectra
of $\zeta$ Oph \citep{Maier2001} and HD 152236 \citep{Galazut2002}
shows similar indications of a spectral shift, however, neither
dataset is of sufficient S/N to conclude that there is a
significant difference between the observed spectrum and the
laboratory assignment. Motivated by this discrepancy, a group
including two of us (M\'A and BJM) revisited the laboratory
spectrum of the 4051 {\AA} band of C$_3$ with higher resolution
than our observations \citep{McCall2003b}.  Indeed, we found 
the $R(0)$ transition to be blue-shifted from the assignment 
of \citet{Gausset1965} and consistent with our astronomical 
spectra. Since $R(0)$ and $P(2)$ share the same upper state, our
measurement of the $R(0)$ transition implies that the 
calculated position of the $P(2)$ line (Table \ref{tbl-C3moldata})
will shift to the blue and overlap $Q(8)$. This is confirmed by
laboratory measurements \citep{McCall2003b}.

\section{Results and Analysis} \label{results}

\subsection{Column Density Comparisons}

%%%%%%%%%%%%%%%%%%%%%%%%%%%%%%%%%%%%%%%%%%%%%%%%%%%%%%%%%%%%%%%%%%%%%%%%%%%%%%%
%%
%%   Observed Properties of C$_3$ and C$_2$ in Various Diffuse Interstellar Clouds
%%
%%%%%%%%%%%%%%%%%%%%%%%%%%%%%%%%%%%%%%%%%%%%%%%%%%%%%%%%%%%%%%%%%%%%%%%%%%%%%%%

\begin{deluxetable*}{ccccccccc}
\tablecolumns{9}
\tabletypesize{\scriptsize}
\tablewidth{0pc}
\tablecaption{
Observed Properties of C$_3$ and C$_2$
\label{tbl-summary}}
\tablehead{
\colhead{Star} &
\colhead{Name} &
\colhead{$N$(C$_3$)} &
\colhead{$T_{low}$(C$_3$)\tablenotemark{a}} &
\colhead{$T_{high}$(C$_3$)\tablenotemark{b}} &
\colhead{$N$(C$_2$)\tablenotemark{c}} &
\colhead{$T$(C$_2$)\tablenotemark{d}} &
\colhead{$n_{coll}$(C$_2$)\tablenotemark{d}} &
\colhead{$N$(C$_2$)/$N$(C$_3$)} \\
 & & \colhead{($\times 10^{12}$ cm$^{-2}$)} &
     \colhead{(K)} &
     \colhead{(K)} &
     \colhead{($\times 10^{14}$ cm$^{-2}$)} &
     \colhead{(K)} &
     \colhead{(cm$^{-3}$)} &
}
\startdata
HD 21483  &             & 4.74 $\pm$ 0.28  &     40 &    560 & 1.10  $\pm$ 0.30  & 13 $\pm$ 5 & 190 $\pm$ 20                                      & 23.2 $\pm$ 6.5   \\
          &             &  2.2 $\pm$ 0.5 \tablenotemark{f}  &        &        &  0.93 \tablenotemark{k} &        &        &                                          \\
HD 23180  & $o$ Per     & 1.51$\pm$ 0.13   &     90 &    470 & 0.32  $\pm$ 0.12 & 60 $\pm$ 20 & 200 $\pm$ 50                                      & 21.2 $\pm$ 8.2   \\
          &             & $<$1.55 \tablenotemark{f}          &        &        & 0.22 $\pm$ 0.03 \tablenotemark{j} &        &                      &                  \\
HD 24398  & $\zeta$ Per & 1.83 $\pm$ 0.35 &     90 &    235 & 0.31  $\pm$ 0.08 & 80 $\pm$ 15 & 460 $\pm$ 150                                      & 17.0 $\pm$ 5.8   \\
          &             &  1.0 \tablenotemark{g}            &        &        & 0.19 $\pm$ 0.03 \tablenotemark{j} &        &                      &                  \\
HD 24534  &  X Per      & 1.88 $\pm$ 0.19 &     81 &    408 & 0.76  $\pm$  0.20 & 44 $\pm$ 5 & 225 $\pm$ 20                                       & 40.4 $\pm$ 11.6  \\
          &             & $\le$1.5 \tablenotemark{f}          &        &        &  0.53 \tablenotemark{l} &        &                              &                  \\
HD 34078  & AE Aur      & 6.56 $\pm$ 0.68 &    171 &    275 & 1.00  $\pm$  0.09 & 120 $\pm$ 10 & $>$500                                           & 15.2 $\pm$ 2.3   \\
          &             & 2.2 $\pm$ 0.5 \tablenotemark{f}   &        &        &  0.58 \tablenotemark{k} &        &                                &                  \\
HD 36371  & $\chi$ Aur  & 0.43 $\pm$ 0.06  &     38 &    554 & 0.29  $\pm$ 0.11 & \nodata & \nodata                                               & 68.1 $\pm$ 27.9  \\
          &             & $<$1.29 \tablenotemark{f}          &        &        &        &        &                                                 &                  \\
HD 53367  &             & 2.21 $\pm$ 0.26  &     40 &    366 & 0.61  $\pm$ 0.16 & \nodata & \nodata                                               & 27.6 $\pm$ 8.1   \\
          &             & $<$2.2 \tablenotemark{f}          &        &        &        &        &                                                 &                  \\
HD 62542  &             & 10.49 $\pm$ 0.53&     75 &    137 & 0.8   $\pm$ 0.2  \tablenotemark{n}  & 36 $\pm$ 15 & 510 $\pm$ 50                    & 7.6  $\pm$ 1.9   \\
HD 148184 & $\chi$ Oph  & 2.50 $\pm$ 0.43 \tablenotemark{i,e} &        &        & 0.43  $\pm$ 0.05 \tablenotemark{j} &        &                   & 17.2 $\pm$ 3.6   \\
HD 149757 & $\zeta$ Oph &  1.6 \tablenotemark{g}            &        &        & 0.25 $\pm$ 0.07 &        &                                        & 16.7 $\pm$ 5.5   \\
          &             & 1.50 $\pm$ 0.29 \tablenotemark{i,e}&        &        & 0.24 $\pm$ 0.03 \tablenotemark{j} &        &                     &                  \\
HD 152236 &             &  1.8 $\pm$ 0.29 \tablenotemark{i,e} &   66 & \nodata & 0.16  $\pm$ 0.06 \tablenotemark{j} & 36 $\pm$ 10 & 300 $\pm$ 100 & 13.3 $\pm$ 2.8   \\
HD 169454 &             & 2.5 $\pm$ 0.66   &     42 & \nodata & 1.60 $\pm$  0.29 & 50 $\pm$ 10 & $>$500                                           & 64.0 $\pm$ 21.8  \\
          &             & 4.3 $\pm$ 0.5 \tablenotemark{f}   &        &        & 0.70  $\pm$ 0.14 \tablenotemark{j} &        &                     &                  \\
HD 179406 & 20 Aql      &  2.0 \tablenotemark{g}            &        &        & 0.82 $\pm$ 0.15 &        &                                        &                  \\
          &             & 1.3 $\pm$ 0.7 \tablenotemark{f}   &        &        &  0.52 \tablenotemark{k} &        &                                &                  \\
HD 204827 &             & 11.51 $\pm$ 0.87 &     43 & \nodata & 4.40 $\pm$  0.29 & 49 $\pm$ 5 & 630 $\pm$ 200                                     & 38.3 $\pm$ 4.1   \\
          &             & 10.4 $\pm$ 0.5 \tablenotemark{f}  &        &        &        &        &        &                                                           \\
HD 210121 &             &  3.79 \tablenotemark{h} &         &        & 1.00 $\pm$ 0.13 &        &        &                                                           \\
          &             & 1.9 $\pm$ 0.5 \tablenotemark{f}    &        &        & 0.65 $\pm$ 0.15 \tablenotemark{m} &        &        &                               \\
\enddata
\tablenotetext{a}{Temperature that best fits low $J\le14$ levels of C$_3$.}
\tablenotetext{b}{Temperature that best fits high $J>14$ levels of C$_3$.}
\tablenotetext{c}{All values from \citet{Thorburn2003} unless otherwise noted 
and normalized to the oscillator strength $f=$1.0$\times$10$^{-3}$.}
\tablenotetext{d}{Temperatures and densities calculated from $N_{\lambda}$ values 
in Table \ref{tbl-C2dat}.}
\tablenotetext{e}{Reported equivalent widths have been converted to column 
densities using oscillator strengths in Table \ref{tbl-C3moldata} and Equation 2. 
Weighted averages are used where multiple measurements are reported for the same $J$
and the sum of the $N_J$ is reported here. Note that total $N$(C$_3$) reported here is 
roughly an order of magnitude larger than the values listed in \citet{Galazut2002},
which do not include the contributions of all observed rotational levels.
}
\tablerefs{(f) \citet{Oka2003}; (g) \citet{Maier2001}; (h) \citet{Roueff2002};
(i) \citet{Galazut2002}; (j) \citet{vD&B1986}; (k) \citet{Federman1994}; 
(l) \citet{Federman1988}; (m) \citet{Gredel1992}; (n) \citet{GvD&B1993}.  }

\tablecomments{Non-detections of C$_3$ in this work (with the upper limit $N$(C$_3$) 
in 10$^{11}$ cm$^{-2}$ 
in parentheses) are:
HD 11415 (1.8),
HD 20041 (4.2),
HD 21398 (3.3),
HD 22951 (2.5),
HD 24912 (1.6),
HD 35149 (2.0),
HD 37022 (2.1),
HD 37061 (3.3),
HD 41117 (1.8),
HD 42087 (2.1),
HD 43384 (2.6),
HD 206267 (3.5),
HD 207198 (6.1),
HD 210839 (6.1).
\citet{Oka2003} measure C$_3$ (with $N$(C$_3$) in 10$^{12}$ cm$^{-2}$ in parentheses) 
toward:
HD 26571 (2.1$\pm$0.6),
HD 27778 (1.2$\pm$0.3),
HD 29647 (4.6$\pm$1.3),
HD 172028 (3.6$\pm$0.6),
HD 203938 ($\le$1.3),
HD 206267 (2.7$\pm$0.4).
}

\end{deluxetable*}

%%%%%%%%%%%%%%%%%%%%%%%%%%%%%%%%%%%%%%%%%%%%%%%%%%%%%%%%%%%%%%%%%%%%%%%%%%%%%%%

Table \ref{tbl-summary} summarizes our measurements of $N$(C$_3$)
along with previous work. In their determination of C$_3$ column
densities from an extensive, low resolution survey,
\citet{Oka2003} measured the equivalent width of the unresolved
$Q$-branch and assumed that it contained half the total intensity
of the C$_3$ band. While this is a good approximation for a thermal
distribution at $T$$<$50~K, we note that this assumption
underestimates the total column density when there is significant
population in higher $J$ levels. Comparisons of $N$(C$_3$) for
HD 204827 and HD 169454 show either agreement within error, or a
slight overestimate by \citet{Oka2003}. Notably, these are the two
clouds for which we measure only a low temperature component in
the excitation profile. On the other hand, AE Aur and HD 21483,
which show pronounced $R$-branch bandheads (Figure
\ref{Bplots1}), are underestimated by \citet{Oka2003} due to a
significant population of C$_3$ in high rotational levels.

Our determination of $N$(C$_3$) for $\zeta$ Per is larger than the
measurements of \citet{Maier2001} due to our higher S/N, which allows
us to measure the populations in higher $J$ levels. The column
density determination for $\zeta$ Oph by \citet{Roueff2002}, using
the data from \citet{Maier2001}, is consistent with our
measurements and highlights the utility of their model for
determining $N$(C$_3$).

\subsection{Correlations}

\begin{figure*}
\epsscale{1} 
\plotone{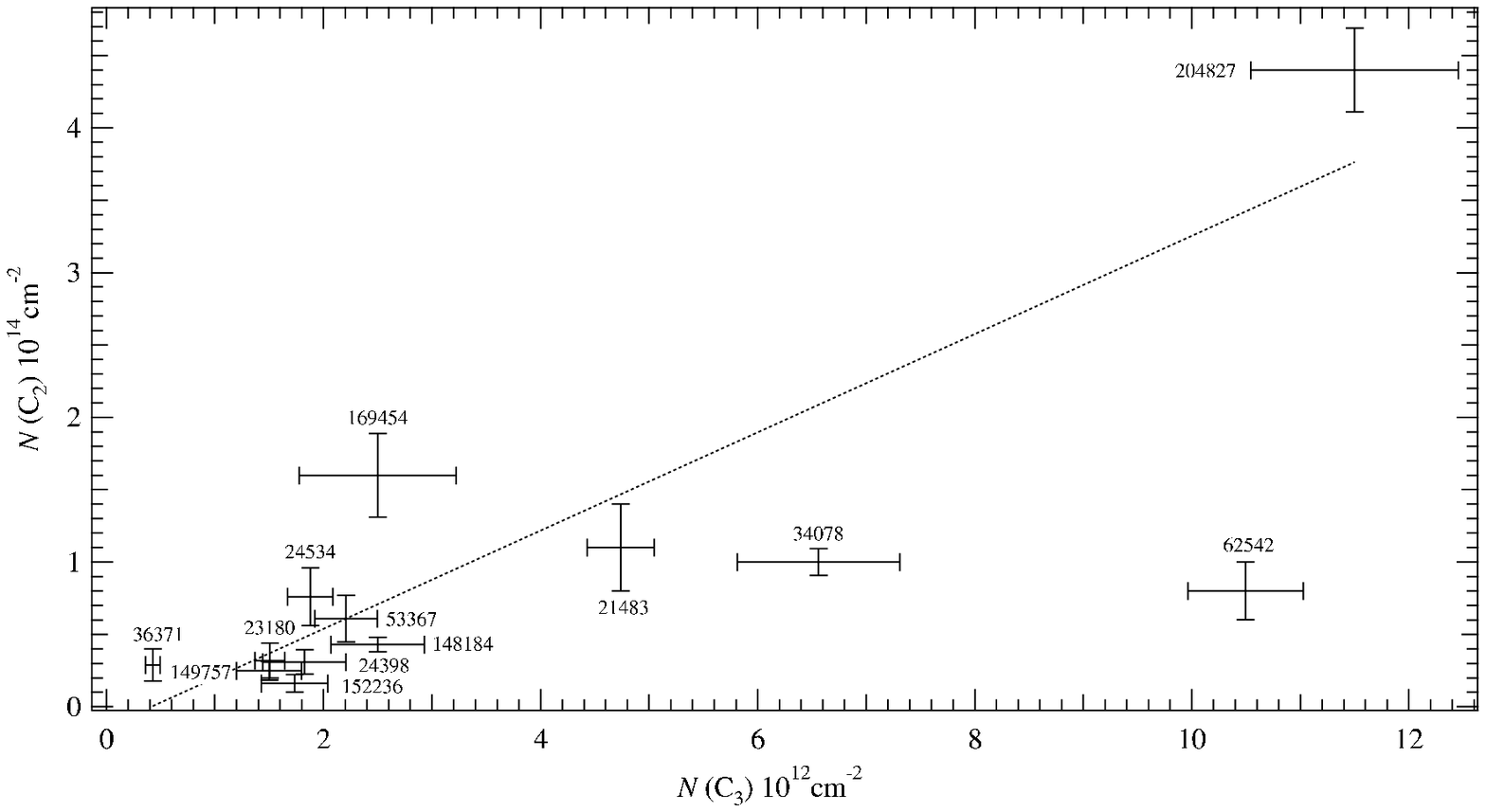} \caption{Comparison of the column
densities of C$_3$ and C$_2$. $N$(C$_3$) determined in this work
is plotted against $N$(C$_2$) determinations of \citet{Thorburn2003}
along with best fit ({\em dotted}) line. Error bars are 1$\sigma$. \label{C2vC3}}
\end{figure*}

The correlation between C$_2$ and C$_3$ pointed out by
\citet{Oka2003} may be tested more rigorously with the precise
determinations of the C$_3$ column density presented here. The
$N$(C$_2$) measurements of \citet{Thorburn2003} are plotted
against our measurements of $N$(C$_3$) in Figure~\ref{C2vC3}. 

The most notable outlier is HD 62542, which has an extremely 
high column density of C$_3$, yet an average to low column of C$_2$.
The intriguing line of sight toward this star seems to be a highly
unorthodox cloud; not only does it seem to be ``missing'' C$_2$, 
it has an extreme UV extinction curve \citep{Cardelli1988},
an unusually low column of CH$^+$ \citep{Cardelli1990}, and it 
lacks the diffuse bands (DIBs) characteristic of reddened 
sightlines \citep{Snow2002}. 
Following the analysis of \citet{Cardelli1990}, \citet{Snow2002} have 
interpreted the lack of DIBs toward this star by suggesting the sight 
line is dominated by a dense cloud stripped of its diffuse outer layers. 
Considering the weak DIBs and the seemingly low column of C$_2$, 
it is noteworthy that a correlation between C$_2$ and some DIBs, 
such as $\lambda$4963 and $\lambda$5769, has recently been identified 
\citep{Thorburn2003}. However, the ``C$_2$ DIBs'' are in general 
much weaker than the classical DIBs, such as $\lambda$6284 and 
$\lambda$5797, which have been reported to be unusually weak 
toward HD 62542 \citep{Snow2002}. Nonetheless, one might speculate 
that the processes which stripped the cloud of its strong DIBs may 
also have destroyed some of its C$_2$. On the other hand,
it is unclear how C$_3$ would survive a process that 
destroys C$_2$, so one could suggest instead that  
the conditions in this cloud may favor carbon chain formation.
Due to the peculiarity of the line of 
sight toward this star, we have excluded it from the correlation 
analysis.

Fitting all of our targets except HD 62542 shows a
correlation between C$_2$ and C$_3$ similar to that
observed by \citet{Oka2003}. The correlation
coefficient for this work is 0.898, whereas \citet{Oka2003}
report 0.932. HD 169454 appears about 2$\sigma$ above the trendline in 
Figure \ref{C2vC3}, but the value of $N$(C$_2$) reported by \citet{Thorburn2003} 
is significantly larger than the values reported by \citet{vD&B1989} and 
\citet{Gredel1986}, which are more consistent 
with the correlation measured here. AE Aur falls significantly 
below the trendline; perhaps the same processes that could have destroyed the 
DIBs and C$_2$ in HD 62542 have occurred, to a lesser extent, in AE Aur 
\citep{Snow2002}. Another consideration is the variability with time of
the molecular species toward AE Aur, where $N$(CH) has increased by 20\% 
over the last 10 years \citep{Rollinde2003}. However, there were 
only 2 years between the observations of C$_2$ and C$_3$ reported here, 
while either $N$(C$_2$) is a factor of $\sim$2 below or $N$(C$_3$) is a factor 
of $\sim$2 too large relative to the general trend observed for the other 
sight lines. The possible enhancement of $N$(C$_3$) is significantly larger 
than expected from the time variability of $N$(CH) alone.

Since HD 204827 has such a large column of C$_3$ relative to the rest of
the sample, much of the correlation seems to rest on this datapoint. 
With the high sensitivity measurements of $N$(C$_3$) presented in this
work, the correlation between C$_2$ and C$_3$ toward stars with relatively
low column densities ($N$(C$_3$)$<$4$\times10^{12}$cm$^{-2}$) may be tested. 
Excluding sources with $N$(C$_3$)$>$4$\times10^{12}$cm$^{-2}$ (i.e. HD 204827, 
HD 62542, AE Aur, and HD 21483) and HD 169454 for reasons discussed above, 
we find a correlation coefficient of 0.841, supporting the correlation determined 
for the entire sample. 

The observed correlation between C$_2$ and C$_3$ is strikingly similar 
to the correlation between C$_2$H and C$_3$H$_2$ measured by \citet{L&L2000}. 
In many cases the ratio of $N$(C$_2$)/$N$(C$_3$) listed in Table 
\ref{tbl-summary} falls within the mean abundance ratio 
$\left<\right.\mathrm{C}_2\mathrm{H}/o$-C$_3\mathrm{H}_2\left.\right>$~= 
27.7$\pm$8 measured by \citet{L&L2000}. 
It is unclear if this is coincidence or if there is some underlying 
chemistry which links these species and their relative abundances.

%%%%%%%%%%%%%%%%%%%%%%%%%%%%%%%%%%%%%%%%%%%%%%%%%%%%%%%%%%%%%%%%%%%%%%%%%%%%%%%
%%
%%   FIGURE: C2 vs C3 Column Density Comparison
%%
%%%%%%%%%%%%%%%%%%%%%%%%%%%%%%%%%%%%%%%%%%%%%%%%%%%%%%%%%%%%%%%%%%%%%%%%%%%%%%%

%%%%%%%%%%%%%%%%%%%%%%%%%%%%%%%%%%%%%%%%%%%%%%%%%%%%%%%%%%%%%%%%%%%%%%%%%%%%%%%
%%
%%   FIGURE: C2 vs C3 Temperature and Density Comparison
%%
%%%%%%%%%%%%%%%%%%%%%%%%%%%%%%%%%%%%%%%%%%%%%%%%%%%%%%%%%%%%%%%%%%%%%%%%%%%%%%%

\begin{figure*}
\epsscale{1}
\plotone{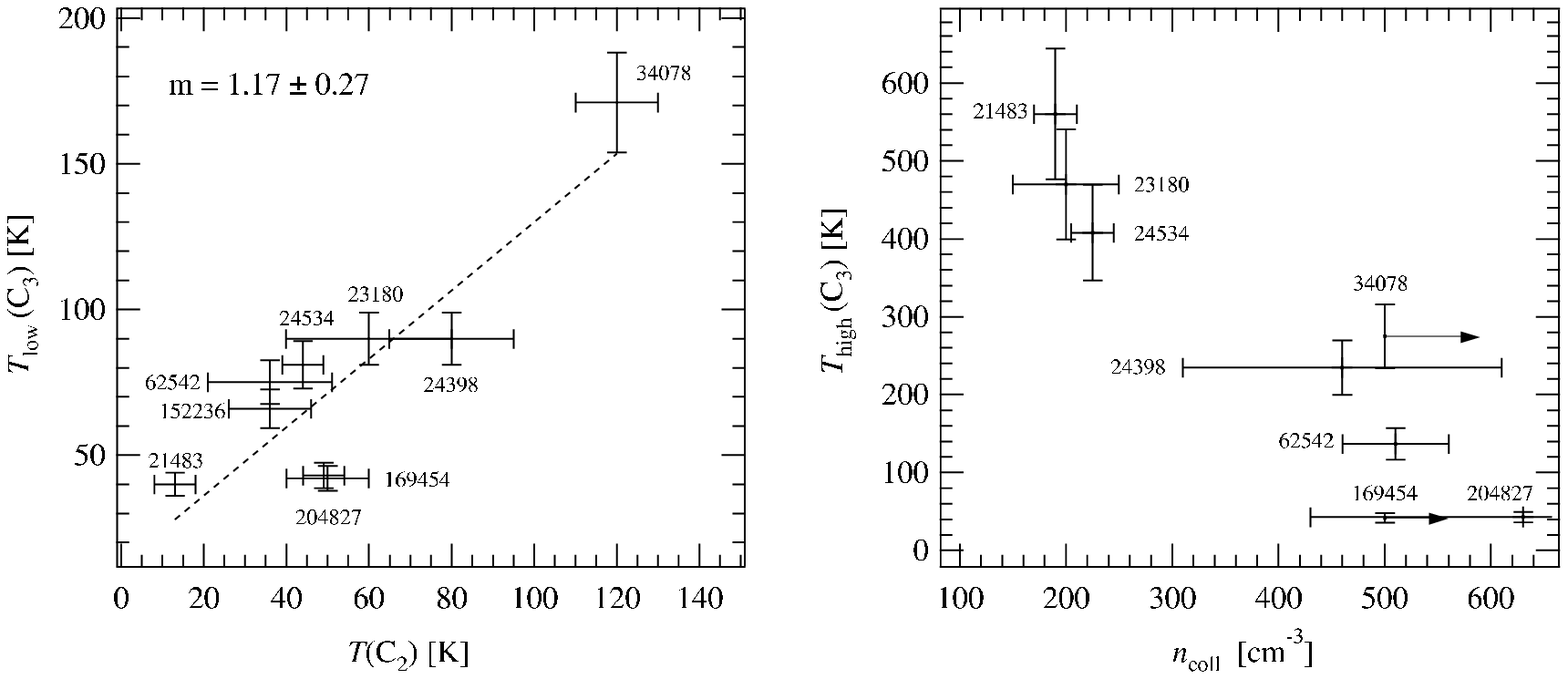} \caption{Comparison of the
excitation of C$_3$ and C$_2$. The excitation temperatures determined
from the observed populations in low ($J$$\le$14) levels of C$_3$ are
compared to the kinetic temperatures determined from the observed
excitation of C$_2$ and full excitation model of \citet{vD&B1986}
({\em left}), showing good agreement. The excitation temperature
determined from the high ($J$$>$14) levels of $C_3$ is compared to the
densities derived from the C$_2$ excitation model ({\em right}).
As expected, T$_{\rm high}$(C$_3$) increases with decreasing
density. \label{temps}}
\end{figure*}

\subsection{The Excitation of C$_2$}

The excitation profile of C$_3$ was compared to the physical
conditions determined by the well-understood rotational excitation
of C$_2$. Table \ref{tbl-C2dat} summarizes previous observations
in the literature of rotationally resolved C$_2$ toward the stars
where we detected C$_3$. To the existing measurements we have added 
our C$_2$ spectrum of HD 204827 measured at Lick and the spectrum 
of X~Per measured at Apache Point Observatory 
(Thorburn 2003, private communication).

%%%%%%%%%%%%%%%%%%%%%%%%%%%%%%%%%%%%%%%%%%%%%%%%%%%%%%%%%%%%%%%%%%%%%%%%%%%%%%%
%%
%%   Observations of Interstellar C2 Towards Various Stars
%%
%%%%%%%%%%%%%%%%%%%%%%%%%%%%%%%%%%%%%%%%%%%%%%%%%%%%%%%%%%%%%%%%%%%%%%%%%%%%%%%

\begin{turnpage}
\begin{deluxetable*}{ccccccccccccc}
%% For lamdscape table in AATex, use rotate command and delete the
%%  RevTeX environment: \begin{turnpage} ... \end{turnpage} environemnt needed 
%%  for EmulateApJ 
%% \rotate
\tablecolumns{13}
\tabletypesize{\scriptsize}
\tablewidth{0pt}
\tablecaption{
Observations of Interstellar C$_2$ Toward Various Stars
\label{tbl-C2dat}}
\tablehead{ & & & \multicolumn{10}{c}{$W_{\lambda}$ (m\AA)} \\
\colhead{$\lambda$ (\AA)} &
\colhead{Line} &
\colhead{$f_{J',J''}$} &
\colhead{HD 21483\tablenotemark{a}}   &
\colhead{HD 23180\tablenotemark{b}}   &
\colhead{HD 24398\tablenotemark{c}}   &
\colhead{HD 24534\tablenotemark{d}}   &
\colhead{HD 34078\tablenotemark{a}}   &
\colhead{HD 62542\tablenotemark{e}}   &
\colhead{HD 149757\tablenotemark{f}}  &
\colhead{HD 152236\tablenotemark{g}}  &
\colhead{HD 169454\tablenotemark{g}}  &
\colhead{HD 204827\tablenotemark{h}}}
\startdata
8757.69 & R(0) & 1.000 & 9.3 $\pm$ 1.2 & 1.3 $\pm$ 0.6 & 0.57 $\pm$ 0.17 & 1.2 $\pm$ 0.2 & \nodata & 4.4 $\pm$ 2 & 0.7 $\pm$ 0.3 & 0.7 $\pm$ 0.5 & 5.6 $\pm$ 1 & 12.8 $\pm$ 5.5 \\
8753.95 & R(2) & 0.400 & 5.1 $\pm$ 1.6 & 1.4 $\pm$ 0.6 & 1.06 $\pm$ 0.17 & 1.9 $\pm$ 0.5 & 4.2 $\pm$ 1 & 9 $\pm$ 1 & 1.6 $\pm$ 0.3 & 1 $\pm$ 0.5 & 6 $\pm$ 0.5 & 18.3 $\pm$ 2.3 \\
8761.19 & Q(2) & 0.500 & 11.5 $\pm$ 1.3 & 1.5 $\pm$ 0.6 & 1.19 $\pm$ 0.17 & 2.8 $\pm$ 0.2 & 3.2 $\pm$ 0.7 & 10 $\pm$ 1 & 1.2 $\pm$ 0.3 & 1.2 $\pm$ 0.5 & 6.6 $\pm$ 0.8 & 23.3 $\pm$ 1.8 \\
8766.03 & P(2) & 0.100 & 5.6 $\pm$ 1.2 & \nodata & \nodata & 0.5 $\pm$ 0.2 & \nodata & 1 $\pm$ 0.5 & \nodata & \nodata & 1.3 $\pm$ 0.3 & 1.4 $\pm$ 3 \\
8751.68 & R(4) & 0.333 & \nodata & \nodata & 0.82 $\pm$ 0.17 & \nodata & 3.4 $\pm$ 0.9 & 4.8 $\pm$ 1.2 & \nodata & \nodata & \nodata & \nodata \\
8763.75 & Q(4)  & 0.500 & 7 $\pm$ 1.3 & 1.7 $\pm$ 0.6 & 1.49 $\pm$ 0.17 & 2.9 $\pm$ 0.4 & 5.3 $\pm$ 0.6 & 7.1 $\pm$ 2 & 1.5 $\pm$ 0.3 & 0.9 $\pm$ 0.5 & 7.9 $\pm$ 0.7 & 15.7 $\pm$ 2.3 \\
8773.43 & P(4) & 0.167 & \nodata & \nodata & \nodata & \nodata & \nodata & 2.6 $\pm$ 0.6 & \nodata & \nodata & 1.7 $\pm$ 0.8 & \nodata \\
8750.85 & R(6)  & 0.308 & \nodata & 1.7 $\pm$ 0.6 & 1.01 $\pm$ 0.17 & 3.2 $\pm$ 0.4 & \nodata & 2.8 $\pm$ 1 & 1.7 $\pm$ 0.3 & \nodata & \nodata & 7.2 $\pm$ 2.6 \\
8767.76 & Q(6) & 0.500 & 8.1 $\pm$ 1.6 & 1.8 $\pm$ 0.6 & \nodata & 2 $\pm$ 0.1 & 4.6 $\pm$ 0.7 & 1.9 $\pm$ 0.9 & 1.4 $\pm$ 0.3 & 1 $\pm$ 0.5 & 3.9 $\pm$ 0.5 & 14.3 $\pm$ 1.1 \\
8782.31 & P(6) & 0.192 & \nodata & \nodata & \nodata & 0.1 $\pm$ 0.3 & \nodata & \nodata & \nodata & \nodata & \nodata & 7.3 $\pm$ 1.7 \\
8751.49 & R(8) & 0.294 & \nodata & \nodata & 0.37 $\pm$ 0.17 & \nodata & \nodata & \nodata & \nodata & 0.4 $\pm$ 0.4 & \nodata & \nodata \\
8773.22 & Q(8) & 0.500 & \nodata & 0.8 $\pm$ 0.6 & \nodata & \nodata & \nodata & 3.2 $\pm$ 0.9 & 1.9 $\pm$ 0.3 & \nodata & 1.5 $\pm$ 0.5 & \nodata \\
8792.65 & P(8) & 0.206 & \nodata & \nodata & \nodata & 3 $\pm$ 0.2 & \nodata & \nodata & \nodata & \nodata & \nodata & \nodata \\
8753.58 & R(10) & 0.286 & \nodata & \nodata & \nodata & 0.7 $\pm$ 0.1 & \nodata & \nodata & 0.4 $\pm$ 0.3 & \nodata & \nodata & \nodata \\
8780.14 & Q(10) & 0.500 & \nodata & \nodata & \nodata & \nodata & \nodata & \nodata & \nodata & \nodata & 1 $\pm$ 0.8 & 2.2 $\pm$ 0.8 \\
   &    &    &    &    &    &    &    &    &    &    &    &    \\
   &  J &    & \multicolumn{10}{c}{$N_J$ ($\times 10^{14}$) }    \\ \cline{2-2}\cline{4-13}
   &    &    &    &    &    &    &    &    &    &    &    &    \\
   &  0 &    & 1.4 $\pm$ 0.2 & 0.19 $\pm$ 0.09 & 0.08 $\pm$ 0.03 & 0.17 $\pm$ 0.03 & \nodata & 0.6 $\pm$ 0.3 & 0.1 $\pm$ 0.04 & 0.1 $\pm$ 0.07 & 0.8 $\pm$ 0.2 & 1.89 $\pm$ 0.81 \\
   &  2 &    & 3.1 $\pm$ 0.3 & 0.47 $\pm$ 0.14 & 0.37 $\pm$ 0.04 & 0.8 $\pm$ 0.06 & 1.1 $\pm$ 0.2 & 2.9 $\pm$ 0.2 & 0.45 $\pm$ 0.07 & 0.36 $\pm$ 0.11 & 2.1 $\pm$ 0.1 & 6.77 $\pm$ 0.4 \\
   &  4 &    & 2.1 $\pm$ 0.4 & 0.5 $\pm$ 0.18 & 0.42 $\pm$ 0.04 & 0.86 $\pm$ 0.1 & 1.6 $\pm$ 0.2 & 2.2 $\pm$ 0.3 & 0.44 $\pm$ 0.09 & 0.26 $\pm$ 0.15 & 2.3 $\pm$ 0.2 & 4.61 $\pm$ 0.7 \\
   &  6 &    & 2.4 $\pm$ 0.5 & 0.61 $\pm$ 0.15 & 0.48 $\pm$ 0.08 & 0.61 $\pm$ 0.04 & 1.4 $\pm$ 0.2 & 0.7 $\pm$ 0.2 & 0.52 $\pm$ 0.08 & 0.29 $\pm$ 0.15 & 1.2 $\pm$ 0.2 & 4.24 $\pm$ 0.3 \\
   &  8 &    & \nodata & 0.24 $\pm$ 0.18 & 0.19 $\pm$ 0.09 & 2.14 $\pm$ 0.13 & \nodata & 0.9 $\pm$ 0.3 & 0.56 $\pm$ 0.09 & 0.1 $\pm$ 0.1 & 0.4 $\pm$ 0.2 & \nodata \\
   & 10 &    & \nodata & \nodata & \nodata & 0.35 $\pm$ 0.08 & \nodata & \nodata & 0.21 $\pm$ 0.15 & \nodata & 0.3 $\pm$ 0.2 & 0.64 $\pm$ 0.2 \\
   &    &    &    &    &    &    &    &    &    &    &    &    \\
   & $n_{coll}$\tablenotemark{i} &    & 190 $\pm$ 20 & 200 $\pm$ 50 & 460 $\pm$ 150 & 225 $\pm$ 20 & $>$500 & 510 $\pm$ 50 & 115 $\pm$ 20 & 300 $\pm$ 100 & $>$500 & 630 $\pm$ 200 \\
   & $T$\tablenotemark{i} &    & 13 $\pm$ 5 & 60 $\pm$ 20 & 80 $\pm$ 15 & 44 $\pm$ 5 & 120 $\pm$ 10 & 36 $\pm$ 15 & 43 $\pm$ 20 & 36 $\pm$ 10 & 50 $\pm$ 10 & 49 $\pm$ 5 \\
\enddata
\tablerefs{(a) \citet{Federman1994}; (b) \citet{Hobbs1981}; (c) \citet{Chaffee1980};
(d) Private communication, Thorburn (2003); (e) \citet{GvD&B1993}; (f) \citet{Hobbs1982};
(g) \citet{vD&B1989}; (h) This work.}
\tablenotetext{i}{Estimated systematic uncertainties.}
\end{deluxetable*}
\end{turnpage}

%%%%%%%%%%%%%%%%%%%%%%%%%%%%%%%%%%%%%%%%%%%%%%%%%%%%%%%%%%%%%%%%%%%%%%%%%%%%%%%

All measured equivalent widths were converted to populations using
the standard relationship in Equation 2. For multiple measurements
of the same $J$ level, weighted averages were used for the
population. The rotational populations were
then simulated using the full C$_2$ excitation model of
\citet{vD&B1982}, probing the density-temperature parameter space
that determines the excitation of C$_2$ while minimizing the
difference between the model and observed values\footnote{A web-based
C$_2$ calculator provided by BJM is available at 
http://dibdata.org/}. 
In this way a kinetic temperature $T$(C$_2$) and a derived density 
of collision partners, $n_{coll}$, was determined for all targets.

As detailed by \citet{vD&B1982}, the derived density $n_{coll}$ is 
dependent on the collision cross-section $\sigma_0$ of C$_2$ with H$_2$ 
or H and a scaling factor $I$ for the standard radiation field, which we 
assume to be 2$\times10^{-16}$ cm$^2$ and 1, respectively. Changes in 
these values will scale $n_{coll}$ accordingly.

Figure \ref{temps} ({\em left}) shows a comparison of the 
temperature $T_{low}$(C$_3$) determined by fitting the low 
($J$$\le$14) rotational levels of C$_3$, with the kinetic temperature 
$T$(C$_2$) determined from the full excitation model of C$_2$. 
While not unexpected, there is an excellent relationship between 
the kinetic temperature and the low $J$ level population of C$_3$. 
One would expect the temperature measured by fitting the higher 
($J$$>$14) rotational levels of C$_3$ to be inversely related to 
density, since as the density increases, collisional de-excitation 
is more efficient at reducing the high rotational level populations. 
This is indeed observed in the comparison of $T_{high}$(C$_3$) versus
$n_{coll}$ and is shown in Figure \ref{temps} ({\em right}).

It is interesting to note that the C$_3$ columns measured toward
HD 204827 and HD 169454, which are the two targets that have only
low temperature components (Figure \ref{Bplots1}) due to their
high densities ($n_{coll}$$>$500 cm$^{-3}$), are also the
targets which deviate most from the trend in the relationship 
between $T_{low}$(C$_3$) and $T$(C$_2$). 
In these high density targets the kinetic temperature derived from 
C$_2$ and the temperature that best fits the excitation of C$_3$ are 
the same, whereas the low density ($n_{coll}$$<$500 cm$^{-3}$)
sources all show a somewhat increased excitation temperature of C$_3$. 
This is because at high densities the excitation profiles of both 
C$_2$ and C$_3$ are predominantly determined by the kinetic temperature, 
whereas at low densities the low $J$ populations, along with the
high-lying levels, have a contribution from radiative pumping. 

\section{Summary} \label{summary}

We have observed rotationally resolved spectra of C$_3$ in 10
translucent sight lines. Using a new method of analysis to
accurately retrieve the individual rotational level populations,
comparisons have been made with the excitation of C$_2$, showing
excellent agreement in the measurement of kinetic temperature.
Trends with density are also consistent with the excitation
expected for a non-polar linear molecule. 

With these measurements we have shown that the correlation between 
$N$(C$_2$) and $N$(C$_3$), as pointed out by \citet{Oka2003}, is 
maintained for sight lines with a range of $N$(C$_3$) from
4.3$\times10^{11}$cm$^{-2}$ to 1.2$\times10^{13}$cm$^{-2}$. 
The star HD 62542 does not fall on the correlation trendline between
$N$(C$_2$) and $N$(C$_3$) and is shown to have either a very large 
column of C$_3$ or depletion in C$_2$, concomitant with the known 
weakness of diffuse band features in its spectrum. Similar processes
may be occurring to a lesser extent in AE Aur. 

Our measurements provide a new challenge for the full C$_3$ excitation 
model presented by \citet{Roueff2002}, which has been heretofore
limited by the quality of observational data.

\acknowledgments
We thank the referee, J.~H.~Black, for helpful suggestions that have
improved the manuscript. We also wish to acknowledge helpful conversations
with L.~M.~Hobbs, T.~Oka, T.~P.~Snow, J.~A.~Thorburn, D.~E.~Welty, 
and D.~G.~York. 

The data presented herein were obtained at the
W.M. Keck Observatory and the University of California
Observatories/Lick Observatory.
The W.M Keck Observatory is operated as a scientific
partnership among the California Institute of Technology, the
University of California and the National Aeronautics and Space
Administration, and was made possible by the generous
financial support of the W.M. Keck Foundation. BJM is supported by 
the Miller Institute for Basic Research in Science. 

% Bibliography

\bibliographystyle{apj}
\bibliography{refs}

\end{document}